\PassOptionsToPackage{dvipsnames*,svgnames,x11names}{xcolor}

\documentclass[sigconf, screen]{acmart}

\setcopyright{none}
\renewcommand\footnotetextcopyrightpermission[1]{}

\usepackage{graphicx}
\usepackage{tikz}
\usepackage{soul}
\usepackage{amsmath}
\usepackage{comment}
\usepackage[normalem]{ulem}
\usepackage[htt]{hyphenat}
\usepackage{braket}
\usepackage[ruled,vlined,linesnumbered]{algorithm2e}
\usepackage{listings}
\usepackage{subcaption}
\usepackage{tabularx}
\usepackage{booktabs}
\usepackage{multirow}
\usepackage{makecell}
\usepackage{hyperref}
\usepackage{tcolorbox}
\usepackage{colortbl}
\usepackage{grumble}
\usepackage{arydshln}

\usepackage[inline]{enumitem}
\setlist{noitemsep,topsep=0pt,parsep=0pt,partopsep=0pt}

\usepackage[subtle]{savetrees}

\usepackage{setspace}
\usepackage[margin=5pt,font={stretch=0.9}]{caption}
\usepackage[margin=8pt,font={small,bf,stretch=0.85}]{caption}

\SetKwProg{Fn}{Procedure}{}{end}
\SetKwBlock{Try}{try}{}
\SetKwBlock{Finally}{finally}{end}

\SetCommentSty{mycomment}

\newcommand{\projecttitle}{QuFrame}

\AtBeginDocument{%
  }

\begin{document}

\title{\projecttitle: Pauli Frame Tracking for Fault-Tolerant Quantum Control}

\author{Aleksandra \'{S}wierkowska}
\affiliation{%
  \institution{Technical University of Munich}
  \city{Munich}
  \country{Germany}}
\email{aleksandra.swierkowska@tum.de}

\author{Ilgaz Er}
\affiliation{%
  \institution{Technical University of Munich}
  \city{Munich}
  \country{Germany}}

\author{Emmanouil Giortamis}
\affiliation{%
  \institution{Technical University of Munich}
  \city{Munich}
  \country{Germany}}

\author{Pramod Bhatotia}
\affiliation{%
  \institution{Technical University of Munich}
  \city{Munich}
  \country{Germany}}

\begin{abstract}
Fault-tolerant quantum computing requires continuous quantum error correction (QEC) to suppress physical noise. While Clifford-only workloads can defer corrections to software post-processing, universal computation breaks this paradigm: executing non-Clifford gates via magic-state gadgets introduces hard, real-time control-flow dependencies on error-corrected measurements. Consequently, the controller must extract syndromes, decode them, and update the Pauli frame in real time, closing the entire QEC loop within a strict $1$\,\textmu s cycle. However, prior proposals on Pauli frame tracking remain largely theoretical, leaving the microarchitectural bottlenecks of tracking, resolving, and synchronizing frame updates across distributed control pipelines unaddressed.

To this end, we present \projecttitle{}, a control microarchitecture for real-time Pauli frame tracking that closes the QEC loop within $1$\,\textmu s while sustaining QPU-rate throughput. By introducing QuFrameISA, we cleanly decouple instruction execution from continuous Pauli frame management. Execution is distributed across per-qubit Instruction Resolution Engines (IREs) that resolve instructions against the frame via a single table lookup, while a Pauli Frame Management Unit enables independent, lock-free concurrent commits. To strictly guarantee the $1$\,\textmu s deadline, \projecttitle{} pairs an asynchronous primary decoder with an on-chip fallback decoder that resolves in $144$\,ns, natively reconciling divergent updates in hardware without degrading logical error rates.
\end{abstract}

\maketitle

\pagestyle{plain}

\section{Introduction}
\label{introduction}

\begin{figure}[t]
    \centering
    \includegraphics[width=0.9\linewidth]{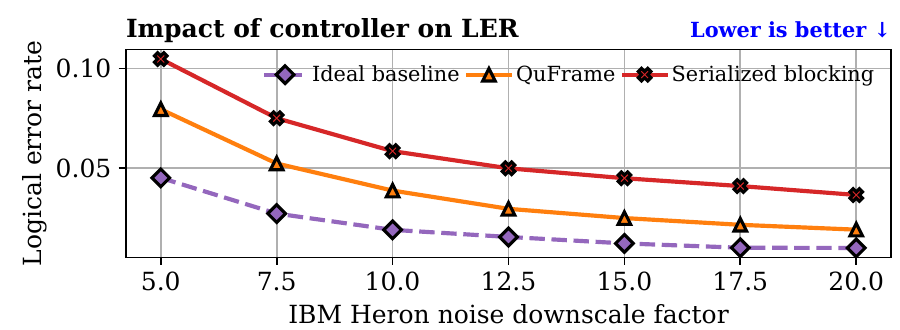}
    \caption{Controller overhead impact on logical error rate (LER). \emph{We evaluate a rotated surface code ($d=3$, 9 rounds, 50,000 shots) on the V80 FPGA, prolonging IBM Heron~\cite{AbuGhanem_2025} coherence times $10\times$ and scaling its noise rates. The ideal baseline assumes no dispatch or decode latency; the blocking baseline serializes frame access and stalls qubits during decoding.}}
    \label{fig:controller_ler}
\end{figure}

Quantum error correction (QEC) is essential for running algorithms on noisy hardware~\cite{Knill_2000}, and universal fault-tolerant computation imposes strict real-time constraints. Non-Clifford operations (such as $T$ gates) rely on magic-state teleportation gadgets whose mid-circuit measurements steer subsequent logical gates~\cite{fowler2012surface}. Consequently, the controller must decode syndromes and react within a single extraction round: approximately $1$\,\textmu s in superconducting architectures~\cite{acharya_2024}. Controllers that fail to meet this deadline accumulate syndrome backlogs that can grow exponentially~\cite{terhal2015quantum}.

Closing this feedback loop requires measuring stabilizers, streaming syndromes to a decoder, and executing the returned corrections~\cite{Nielsen_Chuang_2010, Chatterjee_2023}. Applying every correction as a physical pulse injects unacceptable latency and gate noise. Instead, controllers track a \emph{Pauli frame}: a classical record of accumulated Pauli errors relative to the ideal state~\cite{knill2005, riesebos2017, fu2019control}. Because magic-state gadgets use only Clifford gates and measurements, errors commute through the circuit as Pauli operators, allowing the controller to record corrections purely virtually. The frame defers physical application until a measurement or conditional branch consumes it. Because that measurement requires the prior corrections, frame tracking sits squarely on the critical path, and late decoders stall the entire pipeline.

The Pauli frame must therefore reside directly within the controller's instruction-dispatch datapath. Every gate updates the frame row of its target qubits, and every measurement outcome is corrected against the frame before evaluating branches~\cite{riesebos2017, fu2019control}. Maintaining the frame off-chip (e.g., on an external decoder) would add avoidable round-trip delays to every conditional operation. Real-time frame tracking is fundamentally a controller microarchitecture problem.

Yet no existing controller microarchitecture updates the Pauli frame in real time while supporting multi-core instruction dispatch and absorbing asynchronous decoder results without stalling. Most prior proposals remain conceptual, leaving latency and bandwidth requirements open~\cite{fu2019control, riesebos2017, on2023}. While XQsim synthesizes a control processor with a Pauli frame unit, it drives execution from a single lockstep pipeline that stalls while awaiting corrections~\cite{xqsim}. As we show in Fig.~\ref{fig:controller_ler}, a serialized, stalling controller increases the logical error rate (LER) by up to $4.1\times$ compared to an ideal zero-latency controller, substantially negating physical qubit improvements. This gap leads to our central question:

\statementLeftAccent{Research question}{blue}{\emph{How can a quantum controller track the Pauli frame in real time with bounded per-instruction latency, under multi-core dispatch and decoders of variable latency?}}

\begin{figure*} [t]
    \centering
    \includegraphics[width=\linewidth]{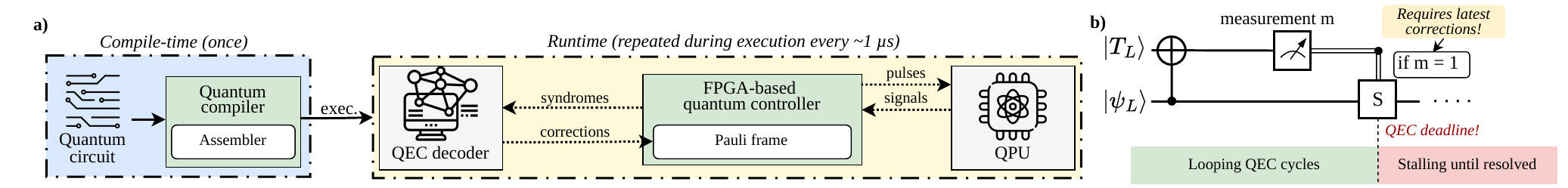}
    \caption{The QEC loop and the conditional gate that sets its deadline. \emph{a) Lifetime of a QEC-protected application, this work's elements in green. b) Magic-state injection and its correction deadline.}}
    \label{fig:background}
\end{figure*}

\myparagraph{Research challenges}
Supporting real-time Pauli frame tracking in hardware introduces three fundamental challenges. \textit{First}, the controller must resolve every outgoing instruction against the frame on the dispatch path, in deterministic, bounded time and without degrading pulse-generation throughput. This cost is paid on every gate: in a rotated surface code, each data qubit takes part in four two-qubit gates per round~\cite{fowler2012surface}. \textit{Second}, state-of-the-art controllers are distributed, multi-core designs driving each physical qubit from an independent execution core~\cite{qubic2023, Yang_2022, fu2018micro, Zhao_2025}, so the frame must be shared state. Even a distance-3 patch spans 17 physical qubits, where each data qubit's entry is accessed by its own core, partner cores during multi-qubit gates, and a decoder every round. Coordinating these concurrent cross-core updates without torn reads or data races risks serializing all 17 cores behind a global lock, turning frame access into a contention bottleneck. \textit{Third}, lookup-table decoders respond within nanoseconds but sacrifice accuracy and scale poorly in memory~\cite{lilliput}. Conversely, high-accuracy decoders exhibit variable turnaround times: Google's real-time decoder averaged $63$\,\textmu s~\cite{acharya_2024}. Stalling that long for a verdict idles qubits for over 50 rounds at feed-forward branch points. To avoid syndrome backlogs, the pipeline must strictly bound both decoding latency and frame updates.

\myparagraph{Key insights}
\textit{(1) Combinational lookups isolate synchronization as the latency
bottleneck.} With two frame bits per qubit and a small native gate set
(seven on IBM Heron~\cite{AbuGhanem_2025}), each core holds a private
read-only copy of the full conjugation table. Instructions resolve in one
combinational lookup, leaving shared-frame access as the sole remaining
latency on the dispatch path.
\textit{(2) Contention is bounded by qubit degree, not patch size.} Each
gate touches at most two rows, and each row is written only by its own
core, its gate partners' cores, and the decoder. Localizing contention allows
per-row versioned compare-and-swap to replace global locks, avoiding stalls on
unrelated qubits with constant-bounded retries.
\textit{(3) Fast and accurate decoders compose without rollback.} Since
Paulis are self-inverse and commute up to phase, a fast, lower-accuracy
decoder can commit immediately while a slower, accurate one repairs
mispredictions later via a single delta conjugated through intervening
Cliffords. Repair applies until the frame is consumed, and even an
unrepaired verdict fails only if its residual error is a nontrivial logical
operator.

\myparagraph{Implementation}
We implement \projecttitle{} as a C++17 reference controller and an HLS-based streaming FPGA microarchitecture synthesized with 17 concurrent cores on an AMD Alveo V80. Each physical qubit pairs with a dedicated core that resolves instructions inline (\emph{Challenge~\#1}) and commits frame updates via per-row compare-and-swap (\emph{Challenge~\#2}). Measurement streams feed a two-tier decoding pipeline in which an on-chip fallback decoder and an external accurate decoder propose corrections in parallel (\emph{Challenge~\#3}).

\myparagraph{Evaluation}
We evaluate \projecttitle{} on a distance-3 rotated surface code under device-calibrated IBM Heron noise models using Helios~\cite{helios}. By eliminating idle stalls and enforcing deterministic frame updates, \projecttitle{} reduces the logical error rate by an average of {$38.5\%$} over a serialized blocking controller (Fig.~\ref{fig:controller_ler}). As physical fidelities improve toward target roadmaps, this reduction reaches {$47.9\%$}, closing {$65.5\%$} of the gap to an idealized zero-latency baseline.

\myparagraph{Key contributions} Our work comprises four core mechanisms:
\begin{itemize}[leftmargin=*]
    \item \textbf{Novel fault-tolerant Instruction Set Architecture (ISA):} We propose \emph{QuFrameISA}, a two-plane ISA that decouples instruction execution from continuous Pauli frame management, accompanied by compiler passes that fix each instruction's schedule and resolution class prior to execution.
    \item \textbf{Ultrafast instruction datapath:} We design the \projecttitle{} datapath around the \emph{Instruction Resolution Engine (IRE)}, which translates every instruction against the live frame with one fixed-cost table lookup, preserving frame correctness for transversal gates.
    \item \textbf{Lock-free synchronization mechanism:} We propose the \emph{Pauli Frame Management Unit}, a versioned, per-row commit protocol that arbitrates concurrent updates from independent cores and asynchronous decoders without any global lock, pairing a driver-follower rule for two-qubit gates with rollback on a failed commit to guarantee exactly-once application.
    \item \textbf{Hierarchical non-stalling decoding:} We introduce the \emph{Decoder Management Unit}, a microarchitectural scheme decoupling asynchronous decoding from real-time execution. Pairing a co-located fallback decoder with a deadline-bound primary solver guarantees corrections within $1$\,\textmu s, while a rolling instruction history buffer lets the pipeline advance with the final verdict in flight.
\end{itemize}

\section{Background}
\label{sec:background}

\begin{figure*}[t]
    \centering
    \vspace{-5px}
    \includegraphics[width=0.9\linewidth]{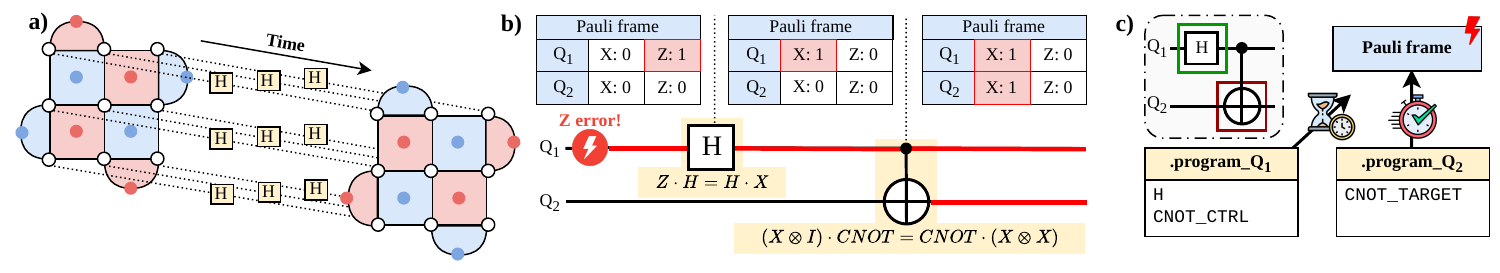}
    \caption{Execution stages of transversal operations. \emph{a) A transversal gate applies to every data qubit; for the rotated surface code the Hadamard also switches stabilizers. b) Virtualized Pauli errors transform through the subcircuits. c) Per-core frame updates can reorder operations.}}
    \label{fig:motivation_fig}
\end{figure*}

In this section, we review the fundamentals of QEC and Pauli frame tracking, examine the execution demands of fault-tolerant workloads, and analyze the microarchitectural bottlenecks in state-of-the-art quantum controllers. 

\subsection{Quantum Error Correction}
\label{subsec:background_qec}

Fault-tolerant quantum computing protects fragile quantum information by encoding \emph{logical qubits} across ensembles of physical qubits~\cite{Nielsen_Chuang_2010}. The QEC loop (Fig.~\ref{fig:background}a) executes continuously: ancilla qubits undergo periodic stabilizer measurements generating syndrome bitstreams, from which a classical decoder infers likely errors and computes corrections for the data qubits~\cite{Chatterjee_2023}. On superconducting architectures, these extraction cycles repeat every $1.1$\,\textmu s throughout a logical qubit's lifetime~\cite{acharya_2024}.

Applying corrective unitary pulses directly to the quantum processor introduces control noise, gate overheads, and accelerates qubit decoherence. To avoid these penalties, architectures virtualize error corrections via Pauli frame tracking~\cite{knill2005, aliferis2008}, where the controller accumulates corrections in classical memory as a two-bit record ($X$ or $Z$ byproducts) per physical qubit. In passive memory experiments, the host can postpone corrections to readout, avoiding real-time decoding backlogs. Prior work also proposed Clifford frames~\cite{Chamberland_2018}, but tracking them demands memory quadratic in the qubit count and places costly matrix operations on the critical path~\cite{Aaronson_2004}. Modern architectures therefore restrict virtualization to the Pauli group, executing Clifford gates directly while updating the frame in real time.

\subsection{Real-Time Demands of Fault-Tolerant Gates}
\label{subsec:background_ft_gates}

Universal logical computation cannot be achieved using transversal operations alone, as dictated by the Eastin-Knill theorem~\cite{Eastin_2009}. In standard 2D surface codes~\cite{fowler2012surface}, transversal non-Clifford gates (such as the $T$ gate) are absent. Fault-tolerant architectures bridge this gap primarily through magic state gadgets or code switching.

Magic state gadgets execute non-Clifford gates indirectly via gate teleportation ~\cite{Bravyi_2005, Bravyi_2012, gidney2024magicstatecultivationgrowing}. As we show in Fig.~\ref{fig:background}b, an injected magic state (e.g., $\ket{T}$) interacts with the logical qubit via transversal Clifford gates and joint stabilizer measurements~\cite{fu2019control}. Because teleportation decomposes entirely into Clifford operations, errors remain within the Pauli group and avoid physical application. However, intermediate syndrome outcomes determine whether the controller must apply a logical $S$ correction. Resolving that branch requires an up-to-date frame, turning offline post-processing into a latency-critical feedback loop.

This dependency places decoding directly onto the execution critical path. Although hardware extracts syndromes every microsecond, software decoders like PyMatching~\cite{higgott2021pymatchingpythonpackagedecoding} struggle to match this throughput. Decoders falling behind the pace of syndrome generation creates an exponential backlog~\cite {terhal2015quantum}, forcing the controller to stall and idle the physical qubits, increasing decoherence and error rates.

Alternatively, code switching transitions logical states between distinct QEC codes with complementary transversal sets. Here, uncorrected Pauli errors would rotate out of the Pauli group into arbitrary continuous rotations under non-Clifford gates, breaking standard Pauli frame tracking~\cite{DiVincenzo2007, Chamberland_2018}. Code switching thus enforces an even stricter deadline: accumulated corrections must be physically applied before switching codes. Although less common, \projecttitle{} also supports physical corrections for full architectural generality.

\subsection{Quantum Controllers}
\label{subsec:background_controllers}

Modern quantum controllers translate compiled circuits into calibrated analog pulses. However, state-of-the-art control hardware remains resource-constrained, typically supporting only 8 to 16 physical qubits per board~\cite{guang2026breakingmemorybottlenecksquantum, Stefanazzi_2022}. While centralized single-core architectures persist~\cite{Stefanazzi_2022}, scalable platforms favor distributed execution~\cite{qubic2023, Yang_2022, lin2019}, pairing each physical qubit with an independent sequencer core. This distributed model provides the execution bandwidth needed to sustain high throughput and prevent serialization delays.

Because traditional distributed controllers handle independent pulse streams without shared execution state, coordination logic focuses almost exclusively on signal-level clock distribution~\cite{qubic2023}. To date, no hardware microarchitecture exists for real-time Pauli frame tracking on a distributed controller, nor does any demonstrate how to resolve the resulting deployment challenges.

\begin{figure*} [t]
    \centering
    \includegraphics[width=0.9\linewidth]{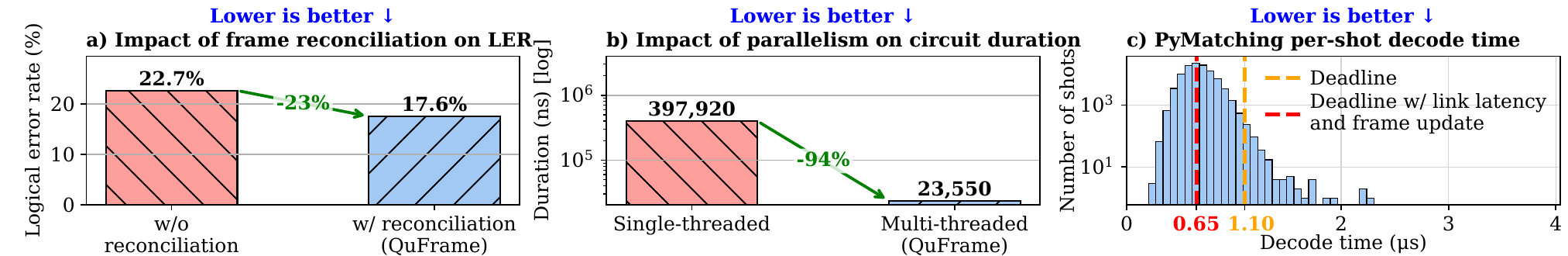}
    \caption{Motivation for the three challenges, evaluated in software on a CPU host. \emph{a) Logical error rate of a transversal-Hadamard distance-5 patch, with and without software Pauli-frame reconciliation (5{,}000 shots). b) Patch execution latency, sequential versus \projecttitle{}'s multi-threaded dispatch. c) Per-shot decode-time distribution over 100{,}000 real PyMatching decodes; dashed lines mark the $1.1\,\mu\text{s}$ round deadline and that deadline minus the real measured CAS-commit and link latency.}}
    \label{fig:motivation}
\end{figure*}

\section{The Case for Real-Time Pauli Frame Tracking}
\label{sec:motivation}
As we discussed in \S\ref{sec:background}, real-time Pauli frame tracking is essential for fault-tolerant execution, yet meeting its stringent latency and correctness bounds introduces three primary architectural challenges.


\subsection{Instruction-Frame Reconciliation}

Before lattice surgery becomes viable for fault-tolerant quantum computation, early quantum processors will primarily rely on transversal operations on limited qubit counts. As Fig.~\ref{fig:motivation_fig}a shows, they apply identical gates across corresponding physical qubits.

Each constituent physical gate transforms the virtual state tracked within the Pauli frame according to its commutation relations (Fig.~\ref{fig:motivation_fig}b). Omitting this tracking causes severe phase and bit-flip misclassifications. As we show in Fig.~\ref{fig:motivation}a, failing to reconcile a single transversal Hadamard gate on a logical qubit increases the logical error rate by 29\%. Because the controller must resolve these updates before dispatching downstream gates, tracking lies directly on the critical path and must complete without stalling the datapath.

\statementLeftAccent{Challenge 1: Real-time frame-instruction reconciliation}{orange}{A virtual error-tracking requires the control architecture to dynamically reconcile Pauli frame updates with in-flight instructions in real time, preventing stalls in the physical datapath.}

\subsection{Synchronization}
The majority of quantum controllers adopt a distributed execution model, dispatching instructions across parallel per-qubit execution lanes to sustain high throughput and avoid decoherence from execution stalls~\cite{qubic2023, Yang_2022, fu2018micro, Zhao_2025}. As we present in Fig.~\ref{fig:motivation}b, serialized execution inflates circuit duration by nearly $17\times$ compared to multi-threaded execution, and idle decoherence degrades the logical error rate, as the serialized blocking scenario in Fig.~\ref{fig:controller_ler} shows.

However, distributed execution introduces data hazards when concurrent execution lanes access shared Pauli frame registers, risking race conditions and out-of-order writes. Because multi-qubit gates require atomic updates across entries for all participating qubits, memory contention persists whether the Pauli frame resides in centralized shared memory or distributed per-core block RAM. Furthermore, the QEC decoder must apply asynchronous frame corrections concurrently with deterministic circuit execution.

Coarse-grained mutual exclusion across the frame table serializes execution and destroys the latency advantages of distributed hardware. Preserving parallel throughput requires a more sophisticated thread-safe synchronization mechanism that resolves concurrent accesses without incurring latency or scalability penalties.

\statementLeftAccent{Challenge 2: Thread-safe frame synchronization}{orange}{The severe synchronization hazards of per-qubit execution demand thread-safe, lock-free microarchitectures to resolve memory contention between independent instruction streams and asynchronous decoders with minimal overhead.}

\subsection{Bounded Latency}


Matching the $1.1$\,\textmu s syndrome generation rate imposes a strict decoding deadline to prevent exponential syndrome backlogs~\cite{acharya_2024, terhal2015quantum}. Yet decoding latency varies widely: algorithms trade accuracy for speed and incur shot-dependent turnaround times~\cite{Higgott_2025}, which Fig.~\ref{fig:motivation}c captures across $100{,}000$ PyMatching shots. Beyond the decoding algorithm, the real-time feedback loop adds substantial system overhead. Updating the Pauli frame consumes controller cycles, while offloading decoding to FPGA accelerators introduces link transport delays~\cite{barber2025realtime, ibm2025qdc, wu2025microblossom}. Recent hardware measurements attribute $77\%$ of a $550$ ns reaction time to classical transport~\cite{yang2026nndecoder}. Modeling this path with Aurora interconnect parameters~\cite{aurora} shows that round-trip transit alone consumes roughly $400$ ns. Combined with the $50$ ns we measure for atomic compare-and-swap (CAS) commits to uncached frame memory, this overhead severely shrinks the remaining compute window, cutting the probability of meeting the $1.1$\,\textmu s deadline from $99.7\%$ to $45.9\%$. This depleted budget forces the controller to execute frame updates nearly instantaneously, or else any added delay directly compounds stalls at feed-forward branches where downstream gates await corrections.

\statementLeftAccent{Challenge 3: Bounded decoding and correcting latency}{orange}{With transport and decoding dominating the microsecond cycle, critical-path frame updates must execute in nanoseconds to prevent feed-forward stalls.}


\subsection{Problem Statement}
While prior control architectures propose general mechanisms for Pauli frame tracking~\cite{riesebos2017, fu2018micro, fu2019control, on2023}, most remain theoretical, omitting physical memory mappings and latency bounds. XQsim~\cite{xqsim} synthesizes a centralized frame unit, but stalls on decoding and restricts execution to rotated surface codes, lattice surgery, and a dedicated decoder. Consequently, existing designs fail to simultaneously guarantee bounded latency, race-free distributed scalability, and generality across codes, decoders, and transversal gates.

To bridge this gap, we require a code- and decoder-agnostic control architecture that seamlessly embeds Pauli tracking directly into the datapath. The resulting microarchitecture must: (i) robustly reconcile frame updates with active instruction streams, (ii) enforce strict thread safety across parallel, per-qubit execution lanes without exacerbating the physical scalability limits of modern control hardware, and (iii) guarantee deterministic, nanosecond-scale latency.
\section{Overview}
\label{sec:overview}
\begin{figure}[t]
    \centering
    \includegraphics[width=0.9\linewidth]{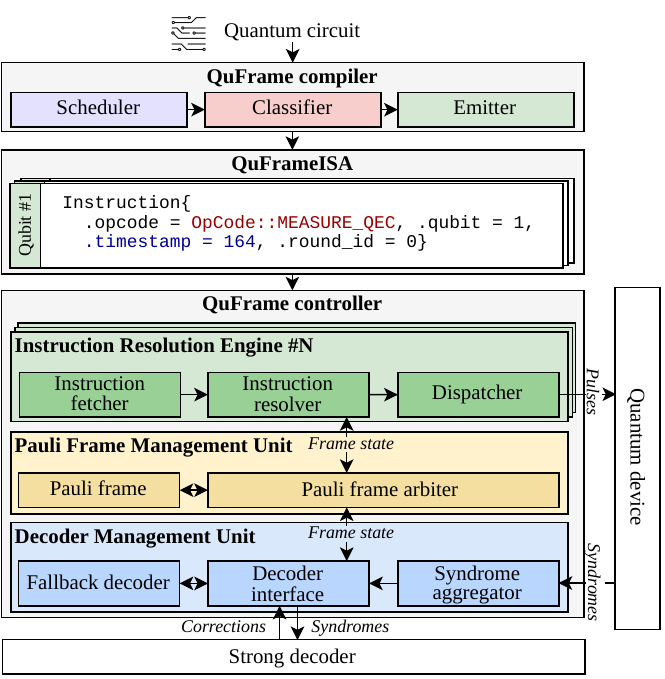}
    \caption{Overview of \projecttitle{}.}
    \label{fig:overview}
\end{figure}
\begin{figure*}[t]
    \centering
    \includegraphics[width=0.77\linewidth]{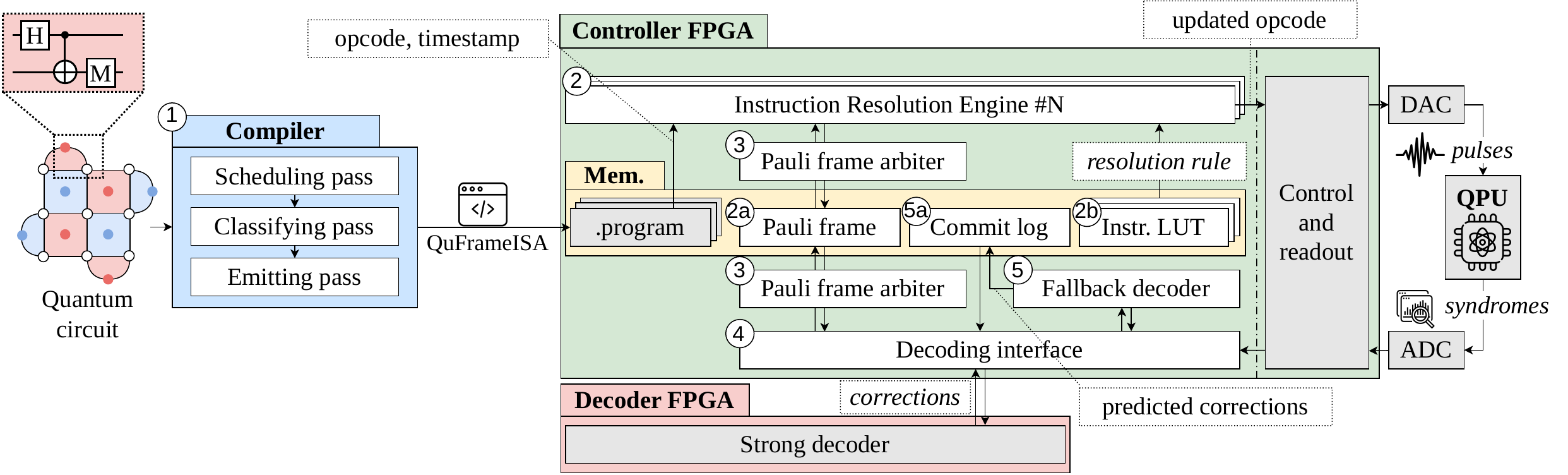}
    \caption{Workflow of \projecttitle{}. \emph{Grey boxes denote the pre-existing modules. The yellow box marks the memory on the controller.}}
    \label{fig:workflow}
\end{figure*}

To address the challenges defined in \S\ref{sec:motivation}, we present \projecttitle{}, an ultra-fast, thread-safe, and code- and decoder-agnostic quantum controller microarchitecture that tracks real-time Pauli frames directly within the hardware datapath. To support independent sequencer cores without cross-core synchronization bottlenecks, we introduce QuFrameISA, a novel instruction set architecture that cleanly decouples physical execution from the frame-tracking control plane.

\subsection{System Components}
As we show in Fig.~\ref{fig:overview}, we design \projecttitle{} as a hardware-accelerated control layer bridging the quantum datapath and the classical decoding pipeline. The architecture consists of four components:

\begin{itemize}[leftmargin=*]
    \item \textbf{Compiler:} A set of compilation passes that schedule operations, classify instructions by frame sensitivity, and emit the circuit as distributed, per-qubit QuFrameISA programs.
    \item \textbf{Instruction Resolution Engine (IRE):} A per-qubit pipeline that executes its instruction stream, resolving each instruction against the Pauli frame via a private read-only conjugation table.
    \item \textbf{Pauli Frame Management Unit:} The shared memory holding the global frame, resolving concurrent IRE and decoder accesses via a versioned, lock-free compare-and-swap.
    \item \textbf{Decoder Management Unit:} A subsystem that aggregates syndromes and dispatches them asynchronously to an external strong decoder, falling back to a co-located fallback decoder's correction whenever the remote solver overruns its deadline.
\end{itemize}

\subsection{QuFrameISA}
\newsavebox{\isacatbox}
\newcommand{\isacat}[1]{\textit{#1}}
\newcommand{\isainst}[1]{\textcolor{red!60!white}{\texttt{#1}}}

\begin{table*}[t]
\centering
\caption{QuFrameISA control-plane primitives.}
\label{tab:isa_extensions}
\footnotesize
\setlength{\tabcolsep}{3pt}
\begin{tabularx}{\textwidth}{l l l X}
\toprule
& \textbf{Instruction} & \textbf{Operands} & \textbf{Description} \\ \midrule
\isacat{Fetch} & \isainst{PROG.NEXT} & \texttt{q} & Return qubit $q$'s next instruction and its scheduled time $t$ \\
\midrule

\multirow{5}{*}{\isacat{Frame}} 
 & \isainst{PFT.READ} & \texttt{q} & Return qubit $q$'s versioned frame row $(x,z,v,t)$ \\
 & \isainst{PFT.CAS} & \texttt{q, v\_exp, x', z', t, h} & Commit qubit $q$'s frame to $(x',z',t)$ if version matches $v_{\text{exp}}$, appending history $h$ for gate commits \\
 & \isainst{PFT.LASTTS} & \texttt{q} & Return qubit $q$'s last committed timestamp only, without its frame bits \\
 & \isainst{PFT.APPLIED} & \texttt{q} & Return qubit $q$'s applied counter \\
 & \isainst{HIST.COMMUTE} & \texttt{delta, q, t\_syn, t\_cur} & Re-derive qubit $q$'s pending correction $\Delta$ through history since $t_{\text{syn}}$ \\
\midrule



\multirow{2}{*}{\isacat{Resolve}} 
 & \isainst{LUT.RESOLVE1/2} & \texttt{frame(s), inst} & Return the updated frame and resolved pulse/outcome for a program-plane opcode \\
 & \isainst{LUT.CONJ1/2} & \texttt{frame, gate} & Pure conjugation function shared by \isainst{LUT.RESOLVE1/2} and \isainst{HIST.COMMUTE} \\
\midrule

\isacat{Sync} & \isainst{SYNC.ARRIVE} & --- & Flip the shared sense bit once every core arrives, releasing waiters \\
\midrule

\isacat{Egress} & \isainst{PULSE.FORWARD} & \texttt{inst} & Send the resolved instruction to the DAC-facing egress queue \\
\midrule

\multirow{2}{*}{\isacat{Dec.}} 
 & \isainst{CLOG.PUSH / CLOG.POP} & \texttt{t\_syn, C / t\_syn} & Record or retrieve the fast decoder's correction $\mathbf{C}$ for syndrome time $t_{\text{syn}}$ \\
 & \isainst{LINK.TX / LINK.RX} & \texttt{syndrome / C, t\_syn} & Send or receive a syndrome, or correction $(\mathbf{C}, t_{\text{syn}})$, over the link to/from the decoder FPGA \\

\bottomrule
\end{tabularx}
\end{table*}

At the core of \projecttitle{} lies QuFrameISA, a novel two-plane ISA for real-time QEC. Following the distributed-core paradigm of quantum controllers, each physical qubit executes an independent, timestamped instruction stream. We decouple the architecture into two planes:

\begin{itemize}[leftmargin=*]
    \item \textbf{The program plane} carries the compiled per-qubit operations: non-Pauli gates emit physical pulses and delegate frame updates to the control plane, whereas Pauli gates outside dynamical-decoupling sequences virtualize into state updates.
    \item \textbf{The control plane} provides the hardware primitives (conjugation lookups, commuting logic, atomic updates) that resolve instructions inline. Decoder corrections map to the same primitives as programmatic Pauli instructions, while lock-free updates to memory ensure safe concurrent execution across cores.
\end{itemize}

\subsection{Workflow Walkthrough}
Fig.~\ref{fig:workflow} illustrates the end-to-end execution workflow:

\circled{1} The compiler lowers the circuit into QuFrameISA in three passes: timestamping gates to synchronize multi-qubit operands and round barriers, assigning instructions to execution classes, and emitting statically scheduled per-qubit programs.

\circled{2} At runtime, each qubit's IRE executes its assigned stream. The IRE reads the current frame state \circled{a} and evaluates the operation against its combinational lookup table \circled{b}. Depending on the class, it absorbs Pauli gates directly into frame updates, conjugates Clifford gates to update the frame prior to dispatch, or issues physical measurement pulses.

\circled{3} The Pauli frame arbiter coordinates updates via a versioned compare-and-swap (CAS) interface to avoid data races. Single-qubit operations perform a single CAS on their local entry. For two-qubit gates, the lower-indexed qubit executes atomic CAS updates across both participating entries while the partner lane stalls.

\circled{4} At round boundaries, the interface aggregates ancilla measurement outcomes and dispatches the resulting syndrome in parallel to both the on-chip fallback decoder and the stronger external decoder.

\circled{5} The fallback decoder completes within nanoseconds, committing a tentative correction to the frame table to maintain execution without stalls. Concurrently, it logs the correction in the commit log \circled{a}, allowing the reconciliation unit to apply the delta once the delayed verdict arrives from the strong decoder.

\begin{figure*}[t]
    \centering
    \vspace{-3px}
    \includegraphics[width=0.85\linewidth]{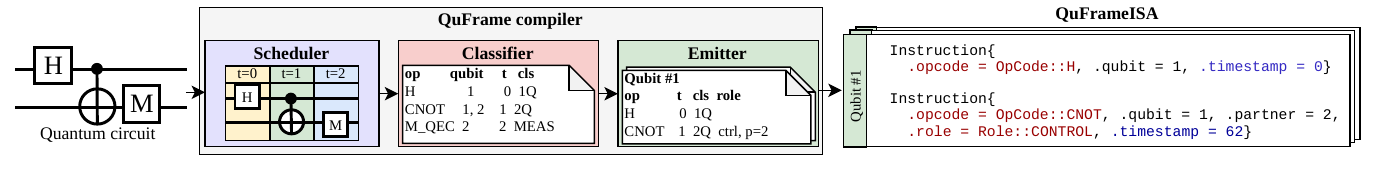}
    \caption{\projecttitle{} compiler. \emph{We first schedule the physical instructions, then classify them according to how they need to be handled by the IRE, and finally emit a program per qubit.}}
    \vspace{-3pt}
    \label{fig:compiler}
\end{figure*}

\subsection{Key Ideas}
We design \projecttitle{} around key design goals that address the fundamental challenges of realizing real-time QEC on the controller (\S\ref{sec:motivation}):

\myparagraph{Key idea \#1 (Challenge 1): \textit{Rule-based datapath resolution}} An IRE sits inline on the datapath, resolving each instruction against a hardcoded conjugation LUT. This fixed-cost lookup ensures every instruction applies its frame update without overrunning its deadline.

\myparagraph{Key idea \#2 (Challenge 2): \textit{Lock-free state arbitration}} The Pauli Frame Management Unit provides a versioned compare-and-swap on each frame row. This localizes contention strictly to concurrent updates on the exact same qubit, ensuring thread safety across asynchronous cores without pipeline stalls.

\myparagraph{Key idea \#3 (Challenge 3): \textit{Decoding fallback}} A co-located fallback decoder injects a fast preliminary correction to sustain continuous execution, moving the strong decoder off the critical path. The controller drops late verdicts to avoid stalls while seamlessly reconciling on-time verdicts against the preliminary update.

\section{QuFrameISA and Compilation}
\label{isa}

\subsection{ISA for Quantum Error Correction} 
While existing pulse-level instruction sets support mid-circuit measurements and basic classical branching, they cannot process continuous, asynchronous decoder feedback or track dynamic corrections in real time. Enabling real-time QEC requires a dedicated ISA that integrates correction bookkeeping directly into the execution flow.

To that end, we propose \textit{QuFrameISA}, an instruction set designed around the Pauli frame. QuFrameISA retains the per-qubit, timestamped streams of distributed controllers, but makes frame resolution an intrinsic execution step: the controller resolves each instruction against its qubit's frame before physical dispatch. A unified set of atomic primitives applies all frame updates, whether initiated by programmed gates or asynchronous decoders, ensuring correctness with minimal latency. This runtime efficiency relies on the \projecttitle{} compiler (\S\ref{isa:compiler}), which lowers global circuits into independent, timestamped per-qubit streams and precomputes all static metadata, restricting the runtime operations to a minimum.

\subsection{QuFrameISA Semantics} 
\textit{QuFrameISA} decouples execution into two planes: the \emph{program plane} represents the compiled quantum circuit as independent, per-qubit streams of timed opcodes with pulse-generation metadata, while the underlying \emph{control plane} dictates how the IRE and Decoder Management Unit translate operations and incoming corrections into Pauli frame updates. 

This separation yields three key architectural benefits: {(1)} Program streams remain decoupled from the frame, ensuring circuit portability and zero real-time tracking overhead. {(2)} Confining frame access to minimal atomic primitives simplifies synchronization logic. {(3)} The IRE and Decoder Management Unit share identical primitives, eliminating redundant hardware paths by processing gate-driven and decoder-driven Pauli updates uniformly.


\myparagraph{Program plane}
Every program-plane instruction falls into one of five classes: \emph{First}, Pauli operations execute virtually by absorbing into the frame, except for compiler-flagged dynamic decoupling sequences~\cite{Das_2021}. \emph{Second}, Clifford gates and resets execute physically after a fixed frame update, conjugating or clearing the frame, respectively. \emph{Third}, non-Clifford gates emit a fixed pulse and fold back a byproduct conditioned on both the pending frame and their outcome. \emph{Fourth}, measurements consume and clear the frame to correct readouts, while syndrome extractions additionally route their resolved outcome to the decoder. \emph{Fifth}, conditionals block on another qubit's resolved outcome, placing real-time frame resolution directly on the critical control-flow path.


Alongside the opcode, every instruction carries static compiler metadata (\S\ref{isa:compiler}): timestamp, resolution class, two-qubit partner data (including the partner's gate index), a frame-sensitive flag, syndrome round ID, and an opaque index for pulse or byproduct tables. Only this index is hardware-specific, keeping QuFrameISA backend-agnostic.

\myparagraph{Control plane}
In Table~\ref{tab:isa_extensions}, we group the control-plane primitives by targeted hardware structures. The IRE uses them to fetch instructions, resolve qubit frames, synchronize across distributed cores, and dispatch operations. Concurrently, the Decoder Management Unit uses them to exchange syndromes and corrections with asynchronous decoders and log early fallback corrections.

Routing all state access through shared atomic primitives creates a single entry point into the frame. Because programmed Pauli gates and dynamic decoder corrections invoke identical conjugation logic, QuFrameISA defines a gate's impact exactly once, ensuring static circuits and dynamic decoders maintain a synchronized frame view.

\begin{figure*}[t]
    \centering
    \vspace{-5pt}    \includegraphics[width=0.85\linewidth]{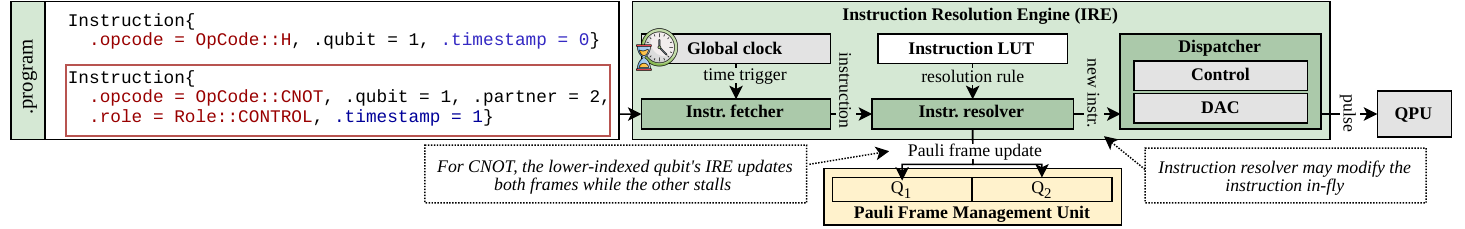}
    \caption{Overview of the Instruction Resolution Engine. \emph{Grey boxes mark pre-existing elements; the white box denotes structure in the memory.}}
    \label{fig:ire}
\end{figure*}

\subsection{\projecttitle{} Compiler}
\label{isa:compiler}
To minimize runtime reconciliation latency, the \projecttitle{} compiler precomputes all frame-dependent instruction metadata before execution, turning a global OpenQASM~2.0 circuit into one independent QuFrameISA stream per physical qubit in three passes (Fig.~\ref{fig:compiler}):

\myparagraph{Scheduler}
The scheduler assigns nanosecond timestamps using as-soon-as-possible list scheduling while preserving per-qubit program order. Each instruction occupies its qubit for its physical gate duration plus the IRE's dispatch latency. Two-qubit gates start only when both operands become available, ensuring identical timestamps across local streams. Barriers align all qubits to the latest timestamp and finalize the syndrome round. In Fig.~\ref{fig:compiler}, the $H$ gate occupies qubit 1 from 0 to 62\,ns, forcing the subsequent $\mathrm{CNOT}$ to start at 62\,ns on both qubits while qubit 2 idles.

\myparagraph{Classifier}
To determine how the IRE resolves each instruction, the classifier assigns each instruction its resolution class, opaque pulse-table index, and frame-sensitive flag. It also tags stabilizer extractions with their current syndrome round and assigns them the $\mathrm{M\_QEC}$ opcode, as we demonstrate in Fig.~\ref{fig:compiler}.

\myparagraph{Emitter}
The emitter generates independent program streams for each physical qubit. It inserts two-qubit operations into both operand programs with matching timestamps and pulse indices, attaching per-qubit context: the partner's identifier and the control or target role for asymmetric gates like $\mathrm{CNOT}$. In Fig.~\ref{fig:compiler}, qubit~1 receives the $H$ gate alongside the control instance of the $\mathrm{CNOT}$ targeting qubit~2.
\section{Instruction Resolution Engine (IRE)}
\label{ire}
Applied gates and asynchronous corrections continuously mutate the Pauli frame, forcing the controller to resolve every instruction against it within a strict nanosecond budget. To prevent datapath stalls, \projecttitle{} equips each physical qubit core with a dedicated IRE. As we illustrate in Fig.~\ref{fig:ire}, the IRE operates as a three-stage pipeline: the \emph{instruction fetcher} retrieves instructions from local memory, the \emph{instruction resolver} updates the frame and modifies the operation via fixed lookup rules, and the \emph{dispatcher} forwards the finalized operation to the digital-to-analog conversion paths. Because each IRE resolves operations via a private read-only lookup table and retries frame commits a bounded number of times, instruction latency remains bounded by a per-class constant, independent of qubit count or circuit depth.

\subsection{Instruction Fetcher}
To ensure physical pulses execute on schedule without idling qubits, the compiler provisions a dispatch timestamp for each instruction that covers gate duration and IRE latency. A \texttt{PFT.CAS} fails only when another writer commits to the same row between the resolver's read and commit attempts, invalidating at most one attempt per interfering write. Because no two IREs write a row concurrently (the follower of a two-qubit gate waits for its driver), the only competing writers are the decoder paths. Since late-correction replay never propagates into ancilla rows (\S\ref{design:non-stalling}), decoders commit to a given row at most $k{=}2$ times per syndrome round (once for fallback correction, once for reconciliation). Consequently, resolution retries at most $k$ times per operand, bounding the cost of an $n$-qubit instruction to its contention-free class latency plus $nk$ retries. The fetcher issues a \texttt{PROG.NEXT} fetch when the global clock matches the instruction timestamp, ensuring resolution finishes before the physical pulse deadline. Hardware barriers and terminations execute entirely within the fetcher, which handles synchronization directly without invoking the resolver. As we show in Fig.~\ref{fig:ire}, qubit 1's fetcher pulls a $\mathrm{CNOT}$ to initiate this process.

\subsection{Instruction Resolver}

\begin{table}[t]
\centering
\caption{Gate resolutions and their corresponding frame updates. \emph{$\mathrm{MEASURE}$ clears the local frame only after execution.}} 
\label{tab:clifford_ire_lut}
\footnotesize
\begin{tabular}{@{} l c c l @{}}
\toprule
\textbf{Initial gate} & \textbf{Initial frame} & \textbf{Updated frame} & \textbf{Updated gate} \\
\midrule
$X$ & $(X_F, Z_F)$ & $(X_F \oplus 1,\, Z_F)$ & $\mathrm{NOP}$ \\
$Y$ & $(X_F, Z_F)$ & $(X_F \oplus 1,\, Z_F \oplus 1)$ & $\mathrm{NOP}$ \\
$Z$ & $(X_F, Z_F)$ & $(X_F,\, Z_F \oplus 1)$ & $\mathrm{NOP}$ \\
\midrule
$H$ & $(X_F, Z_F)$ & $(Z_F, X_F)$ & $H$ \\
$S, S^\dagger$ & $(X_F, Z_F)$ & $(X_F,\, Z_F \oplus X_F)$ & $S/S^\dagger$ \\
\midrule
$\mathrm{CNOT}(A,B)$ & \makecell{$(X_A, Z_A)$ \\ $\otimes (X_B, Z_B)$} & \makecell{$(X_A,\, Z_A \oplus Z_B)$ \\ $\otimes (X_B \oplus X_A,\, Z_B)$} & $\mathrm{CNOT}(A,B)$ \\
\midrule
$\mathrm{MEASURE}$ & $(X_F, Z_F)$ & $(0,0)$ & $\mathrm{MEASURE}$ \\
\bottomrule
\end{tabular}
\end{table}

The resolver forms the core of the \projecttitle{} architecture. It evaluates each incoming instruction against the active Pauli frame to produce an updated frame state and a final hardware instruction. As we detail in Table~\ref{tab:clifford_ire_lut}, the resolver tailors these outputs to the instruction class: it absorbs Pauli gates into the frame while suppressing the physical instruction, and forwards non-Pauli Clifford gates and measurements unchanged while updating or consuming the frame accordingly.

Upon receiving an instruction, the resolver reads the active frame bits for the target operands. Because resolution rules depend solely on the opcode and two frame bits per operand, the resolver evaluates outcomes through a private, read-only lookup table (LUT) that eliminates cross-core contention. This fixed-cost lookup scales exclusively with instruction-set size, remaining independent of qubit count or circuit depth. The resolver then passes the updated state to the Pauli Frame Management Unit (\S\ref{design:sync}) for atomic commit and forwards physical operations to the dispatcher.

As we illustrate in Fig.~\ref{fig:ire}, a two-qubit $\mathrm{CNOT}$ propagates the control qubit's $X$ error to the target and the target's $Z$ error to the control, matching the rule in Table~\ref{tab:clifford_ire_lut} and Fig.~\ref{fig:motivation_fig}. For cross-core coordination, \projecttitle{} assigns the driver role to the lower-indexed operand. Because qubit 1 holds the lower index, it issues the joint error updates to the Pauli Frame Management Unit (\S\ref{design:sync}), while qubit 2 holds its pipeline until the update completes.

\subsection{Dispatcher}
The dispatcher buffers the resolved instruction until its scheduled dispatch window, then triggers the control logic and Digital-to-Analog Converters (DACs) to generate the physical waveform. Because standard hardware stacks like QubiC~\cite{qubic2023} already provide this pulse-generation path, our evaluation abstracts the physical dispatcher and forwards resolved instructions directly to a simulated QPU.
\section{Decoder Management Unit}
\label{decoder}

\begin{figure}[t]
    \centering
    \includegraphics[width=\linewidth]{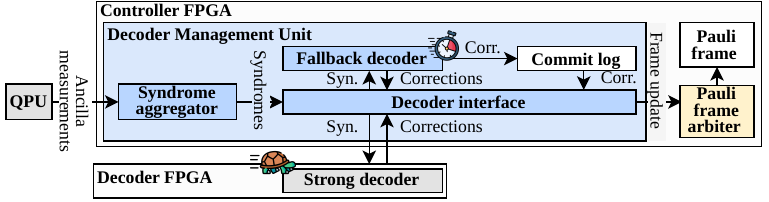}
    \caption{The Decoder Management Unit. \emph{Aggregated syndromes reach a fast fallback decoder and a slow strong decoder hosted externally at once. \projecttitle{} applies and logs the fallback correction immediately, then reverts it if the strong verdict arrives in time.}}
    \label{fig:decoder}
\end{figure}

Executing fault-tolerant quantum algorithms requires rapidly extracting syndromes, routing them to a decoder, and updating the Pauli frame. High-accuracy decoders demand substantial resources, prompting architectures to offload them to external accelerators like FPGAs~\cite{Kurman_2026}. However, when strong decoding and inter-FPGA communication exceed the 1.1 \,\textmu s deadline (\S\ref{sec:background}), the controller must either resolve subsequent instructions against a stale frame or pause execution, which incurs exponentially growing decoherence errors~\cite{holmes2020nisqboostingquantumcomputing}. As we present in Fig.~\ref{fig:decoder}, \projecttitle{} introduces the Decoder Management Unit to resolve this bottleneck. This unit deploys a multi-layer decoding hierarchy that pairs a lightweight fallback decoder for simple, frequent faults with an accurate remote decoder, guaranteeing the controller meets strict execution deadlines~\cite{delfosse2020hierarchicaldecodingreducehardware, Knapen_2026, wegmann2026zerogpredecoderawaredecoderquantum}.

\subsection{Dual-Decoding Path for Guaranteed Latency}
\label{design:fallback-decoder}

Assuming an external accelerator hosts the strong decoder, the Decoder Management Unit coordinates three dedicated modules:

\myparagraph{Syndrome aggregator} Because each IRE runs independently, measurement outcomes for a single stabilizer round arrive asynchronously. The aggregator groups these incoming outcomes by their compiler-assigned round numbers, without requiring knowledge of the generating qubits. Once received outcomes match the compiler-registered count, the aggregator submits the complete syndrome to both decoding paths via the decoder interface.

\myparagraph{Fallback decoder}
We implement the fallback path as an on-chip LUT decoder inspired by LILLIPUT~\cite{lilliput}, reducing decoding to a single direct-indexed memory read with fixed latency. Indexed by current-round syndrome bits, the table logs and forwards precomputed corrections for all error patterns up to weight~6 to the decoder interface. At distance-3, this enumeration is exhaustive and resolves every round, whereas larger codes bound the error weight for tractability. Because low-weight events account for over 90\% of physical errors~\cite{wegmann2026zerogpredecoderawaredecoderquantum}, table misses remain rare. On a miss, the LUT issues a null correction and defers the round to the strong decoder.

\myparagraph{Decoder interface} The decoder interface manages dual-path arbitration while keeping the execution pipeline agnostic to strong decoder implementations. If a strong verdict misses its timeout cutoff, the interface drops it to avoid pipeline stalls and retains the fallback's preliminary correction. When an on-time strong verdict arrives, the interface retrieves the fallback correction from the commit log and applies only the Pauli difference between the two verdicts. Because Pauli operators are self-inverting, applying this difference reverts the tentative correction and commits the final correction without dedicated recovery logic. Both decode paths access the frame table through the same versioned-commit discipline as local IREs (\S\ref{design:sync}), ensuring the interface acts as a standard, underprivileged client.

\begin{figure*}[t]
\centering
\resizebox{\textwidth}{!}{%
\begin{tikzpicture}[>=stealth, font=\footnotesize,
  lane/.style={->, thick, gray!60},
  bar/.style={draw=gray!50, fill=gray!12},
  op/.style={draw=#1, fill=#1!15, rounded corners=1.5pt, minimum height=0.4cm, inner sep=1.5pt, font=\footnotesize},
  st/.style={draw=gray!60, fill=white, rounded corners=1pt, inner sep=1pt, font=\footnotesize, minimum width=0.62cm, text height=1.5ex, text depth=0.35ex, align=center},
  bad/.style={draw=FireBrick, fill=FireBrick!12, rounded corners=1pt, inner sep=1pt, font=\footnotesize\bfseries, text=FireBrick, minimum width=0.62cm, text height=1.5ex, text depth=0.35ex, align=center},
  rd/.style={->, densely dashed, #1},
  wr/.style={->, thick, #1},
  vl/.style={font=\scriptsize, inner sep=1pt},
  lbl/.style={font=\footnotesize, align=right, anchor=east},
  ttl/.style={font=\small\bfseries, anchor=south, text width=5.3cm, align=center}]

\begin{scope}
  \node[ttl] at (3.25,2.05) {(a) Two IREs overwrite each other};
  \draw[lane] (1.2,1.7) -- (5.3,1.7);  \node[lbl] at (1.15,1.7) {IRE$_1$};
  \draw[bar] (1.2,0.68) rectangle (5.3,1.02); \node[lbl, align=center] at (1.15,0.85) {Frame\\[-2pt]$(Q_1,Q_2)$};
  \draw[lane] (1.2,0) -- (5.3,0);      \node[lbl] at (1.15,0) {IRE$_2$};
  \node[st] at (1.62,0.85) {$X,I$};
  \node[op=ForestGreen, minimum width=0.7cm] at (2.25,1.7) {CNOT};
  \draw[rd=ForestGreen] (1.8,1.02) -- (1.95,1.5) node[vl, midway, left] {$X,I$};
  \draw[wr=ForestGreen] (2.55,1.5) -- (2.7,1.02) node[vl, midway, right] {$Q_2{\leftarrow}X$};
  \node[st] at (3.1,0.85) {$X,X$};
  \node[op=ForestGreen, minimum width=0.7cm] at (3.75,0) {CNOT};
  \draw[rd=ForestGreen] (3.3,0.68) -- (3.45,0.2) node[vl, midway, left] {$X,X$};
  \draw[wr=ForestGreen] (4.05,0.2) -- (4.2,0.68) node[vl, midway, right] {$Q_2{\leftarrow}I$};
  \node[bad] at (4.85,0.85) {$X,I$};
  \node[vl, FireBrick] at (4.85,1.2) {expected $X,X$};
\end{scope}

\begin{scope}[xshift=5.6cm]
  \node[ttl] at (3.25,2.05) {(b) H resolves after the CNOT};
  \draw[lane] (1.2,1.7) -- (5.3,1.7);  \node[lbl] at (1.15,1.7) {IRE$_1$};
  \draw[bar] (1.2,0.68) rectangle (5.3,1.02); \node[lbl, align=center] at (1.15,0.85) {Frame\\[-2pt]$(Q_1,Q_2)$};
  \draw[lane] (1.2,0) -- (5.3,0);      \node[lbl] at (1.15,0) {IRE$_2$};
  \node[st] at (1.62,0.85) {$Z,I$};
  \node[op=ForestGreen, minimum width=0.7cm] at (2.25,0) {CNOT};
  \draw[rd=ForestGreen] (1.8,0.68) -- (1.95,0.2) node[vl, midway, left] {$Z,I$};
  \draw[wr=ForestGreen] (2.55,0.2) -- (2.7,0.68) node[vl, midway, right] {$Z,I$};
  \node[st] at (3.1,0.85) {$Z,I$};
  \node[op=ForestGreen, minimum width=0.7cm] at (3.95,1.7) {H};
  \draw[rd=ForestGreen] (3.5,1.02) -- (3.65,1.5) node[vl, midway, left] {$Z$};
  \draw[wr=ForestGreen] (4.25,1.5) -- (4.4,1.02) node[vl, midway, right] {$Q_1{\leftarrow}X$};
  \node[bad] at (4.85,0.85) {$X,I$};
  \node[vl, FireBrick] at (4.9,0.5) {expected $X,X$};
\end{scope}

\begin{scope}[xshift=11.2cm]
  \node[ttl] at (3.25,2.05) {(c) Stale read after a decoder update};
  \draw[lane] (1.2,1.7) -- (5.3,1.7);  \node[lbl] at (1.15,1.7) {IRE$_1$};
  \draw[bar] (1.2,0.68) rectangle (5.3,1.02); \node[lbl, align=center] at (1.15,0.85) {Frame\\[-2pt]$(Q_1,Q_2)$};
  \draw[lane] (1.2,0) -- (5.3,0);      \node[lbl] at (1.15,0) {Decoder};
  \node[st] at (1.62,0.85) {$Z,I$};
  \node[op=ForestGreen, minimum width=2.4cm] at (3.1,1.7) {H};
  \draw[rd=ForestGreen] (1.8,1.02) -- (1.95,1.5) node[vl, midway, left] {$Z$};
  \node[op=SteelBlue, minimum width=0.7cm] at (2.75,0) {corr.\ $X$};
  \draw[wr=SteelBlue] (3.05,0.2) -- (3.2,0.68) node[vl, midway, right] {$Q_1{\leftarrow}Y$};
  \node[st] at (3.72,0.85) {$Y,I$};
  \draw[wr=ForestGreen] (4.25,1.5) -- (4.4,1.02) node[vl, midway, right] {$Q_1{\leftarrow}X$};
  \node[bad] at (4.85,0.85) {$X,I$};
  \node[vl, FireBrick] at (4.9,0.5) {expected $Y,I$};
\end{scope}

\begin{scope}[xshift=16.8cm]
  \node[ttl] at (3.25,2.05) {(d) CNOT sees a half-applied correction};
  \draw[lane] (1.2,1.7) -- (5.3,1.7);  \node[lbl] at (1.15,1.7) {IRE$_1$};
  \draw[bar] (1.2,0.68) rectangle (5.3,1.02); \node[lbl, align=center] at (1.15,0.85) {Frame\\[-2pt]$(Q_1,Q_2)$};
  \draw[lane] (1.2,0) -- (5.3,0);      \node[lbl] at (1.15,0) {Decoder};
  \node[st] at (1.55,0.85) {$I,I$};
  \node[op=SteelBlue, minimum width=0.7cm] at (1.7,0) {corr.\ $X$};
  \draw[wr=SteelBlue] (1.9,0.2) -- (2.05,0.68) node[vl, midway, right] {$Q_1{\leftarrow}X$};
  \node[st] at (2.6,0.85) {$X,I$};
  \draw[rd=ForestGreen] (2.95,1.02) -- (3.1,1.5) node[vl, midway, left] {$X,I$};
  \node[op=ForestGreen, minimum width=0.7cm] at (3.45,1.7) {CNOT};
  \draw[wr=ForestGreen] (3.75,1.5) -- (3.9,1.02) node[vl, midway, left] {$X,X$};
  \node[st] at (4.25,0.85) {$X,X$};
  \node[op=SteelBlue, minimum width=0.7cm] at (4.3,0) {corr.\ $Z$};
  \draw[wr=SteelBlue] (4.5,0.2) -- (4.65,0.68) node[vl, midway, left] {$Q_2{\leftarrow}Y$};
  \node[bad] at (5.0,0.85) {$X,Y$};
  \node[vl, FireBrick] at (4.75,1.2) {expected $Y,Y$};
\end{scope}

\end{tikzpicture}}
\caption{Synchronization hazards in the program of Fig.~\ref{fig:ire}. \emph{(a) Both IREs apply the CNOT, so the propagated $X$ doubles up and cancels. (b) The CNOT commits before the H. (c) A stale read erases the decoder's correction. (d) The correction reaches $Q_1$ but not $Q_2$.}}
\label{fig:sync_hazards}
\end{figure*}
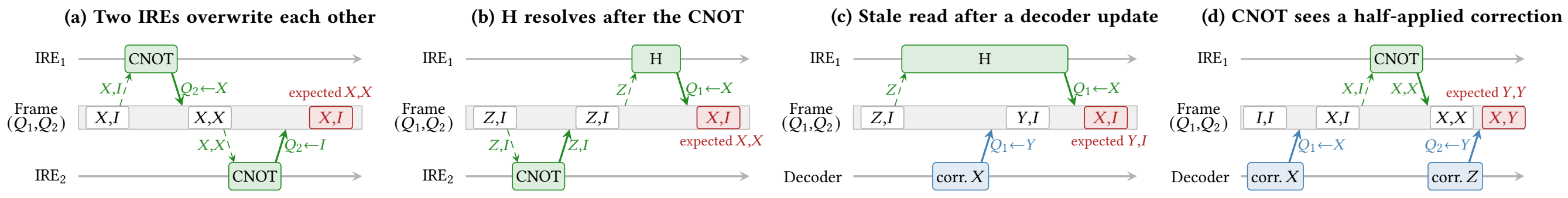

\subsection{Non-Stalling Mechanism}
\label{design:non-stalling}
Stalling execution until a decoder verdict arrives leaves qubits idle, accelerating physical error accumulation through decoherence~\cite{ghosh_2012, Maurya_2025}. To sustain execution, we propose a non-stalling mechanism that optionally appends a fixed-depth \emph{rolling history} to each frame table row (\S\ref{design:sync}). By trading modest per-qubit memory overhead, this history allows the pipeline to advance without delay, applying late corrections asynchronously by commuting them forward (e.g., propagating an $X$ correction through an intermediate $H$ gate into a $Z$). Only instructions requiring a settled frame stall dispatch and bound history traversal: non-Clifford gates fix their byproduct against the frame, and feed-forward conditionals block until their target outcome resolves. To further suppress decoherence, the compiler distributes non-stalling operations evenly across the decoder waiting window rather than clustering them at its edges.

To keep this reconciliation bounded and semantically exact, we enforce four rules: \textit{First}, a pre-measurement correction arriving late patches the recorded classical readout rather than the cleared row. \textit{Second}, a pre-reset correction arriving late is dropped, since the row now tracks a freshly prepared state. \textit{Third}, across an intervening two-qubit gate, the correction propagates to the partner qubit via standard IRE rules. \textit{Fourth}, traversal halts at non-Clifford gates and conditionals, which bind an operation to the frame or a classical outcome.
\section{Pauli Frame Management Unit}
\label{design:sync}

Decentralized execution provides high throughput but introduces severe synchronization hazards when multiple cores update the Pauli frame concurrently (\S\ref{sec:background}). To guarantee data consistency without global locks, \projecttitle{} introduces the Pauli Frame Management Unit. This unit maintains the active Pauli frame for every physical qubit and coordinates thread-safe access, allowing distributed IREs (\S\ref{ire}) and the Decoder Management Unit (\S\ref{decoder}) to read and update states simultaneously without datapath stalls.

\begin{table}[t]
\centering
\caption{Pauli Frame Table after the example of Fig.~\ref{fig:ire}. \emph{Each row commits frame bits, version and timestamp as one atomic word, plus a rolling history and an applied-instruction counter.}}
\label{tab:frame_example}
\footnotesize
\setlength{\tabcolsep}{3pt}
\begin{tabular}{@{} c c c c c >{\raggedright\arraybackslash}p{3.5cm} @{}}
\toprule
\textbf{Qubit} & \textbf{Frame} & \textbf{Version} & \textbf{Time} & {\textbf{Counter}} & \textbf{Rolling history} \\
\midrule
$Q_1$ & X=1, Z=0 & 3 & 62\,ns & {2} & $H$ (t=0, v=2), $\mathrm{CNOT}_{\text{ctrl}}$ (t=62, v=3) \\
$Q_2$ & X=1, Z=0 & 1 & 62\,ns & {1} & $\mathrm{CNOT}_{\text{targ}}$ (t=62, v=1) \\
\bottomrule
\end{tabular}
\end{table}

\subsection{Pauli Frame Structure}
To avoid inter-core messaging, we map the Pauli frame into shared memory with a dedicated row per physical qubit. Per-data-qubit tracking eliminates atomic CAS contention across parallel cores and maintains the fine-grained state needed to flush virtual errors during code switching. Furthermore, tracking ancillas prevents divergent synchronization paths in the IRE, bounding dispatch latency. As we show in Table~\ref{tab:frame_example}, each row contains the \emph{qubit ID} and the active \emph{frame bits} ($X$ and $Z$, tracking virtual errors). In addition, each row stores four metadata fields: a \emph{timestamp} of the latest absorbed instruction, a monotonic \emph{version} tag uniquely identifying the state, an \emph{applied counter} tracking completed local program instructions, and a fixed-depth \emph{rolling history} (\S\ref{design:non-stalling}) logging recent operations. Every commit atomically updates the frame bits, version, timestamp, and history, ensuring readers never observe fractured state.

\subsection{Pauli Frame Synchronization Hazards}
To prevent hardware bottlenecks, \projecttitle{} replaces global locks with decentralized synchronization at strict per-row granularity. However, concurrent multi-core updates introduce critical synchronization hazards:
\textbf{Problem~\#1:} If both participating cores independently apply a two-qubit gate, the propagated error doubles up and cancels, as we illustrate in Fig.~\ref{fig:sync_hazards}a.
\textbf{Problem~\#2:} Decentralized execution can cause out-of-order commits when a driver commits a joint gate to its partner's row before the partner commits an earlier local operation, as we show in Fig.~\ref{fig:sync_hazards}b.
\textbf{Problem~\#3:} If an IRE computes a gate using a stale read, its subsequent commit could silently erase a concurrent write that landed in the interim, such as a decoder correction, as we depict in Fig.~\ref{fig:sync_hazards}c. 
\textbf{Problem~\#4:} Because rows update independently, a decoder correction landing on only one operand prior to a two-qubit gate causes the joint operation to propagate an inconsistent error state, as we present in Fig.~\ref{fig:sync_hazards}d.

\subsection{\projecttitle{} Synchronization Protocol}

To resolve these hazards without global locks, \projecttitle{} combines versioned CAS with a driver-follower protocol. For single-qubit instructions, an IRE executes an optimistic read-compute-update loop: it reads the row, computes the update locally, and commits via CAS, succeeding only if the row's version is unchanged.

Building on this foundation, we resolve the identified hazards:
\textbf{Solution~\#1:} To prevent double application, we enforce a driver-follower protocol for two-qubit gates. Only the lower-indexed core (the \emph{driver}) commits the joint gate to both rows, while the higher-indexed core (the \emph{follower}) idles until its row updates. If the driver's second CAS fails, it rolls back the first commit by inverting the applied delta and retries, ensuring exactly-once application without deadlocks.
\textbf{Solution~\#2:} To prevent out-of-order commits, the protocol uses row applied counters as program-order barriers. The driver commits a joint gate only after the partner's row counter matches that gate's index in the partner program, forcing the follower to complete earlier local operations first.
\textbf{Solution~\#3:} Versioned CAS eliminates stale-read hazards across all writers (IREs, fallback decoders, and remote decoders). Stale versions cause CAS rejection and trigger clean retries, preventing lost updates.
\textbf{Solution~\#4:} The driver-follower protocol prevents fractured error propagation by resolving the joint state before committing to either row. Any intervening correction on one row triggers a CAS failure and a full joint-gate retry.

Ultimately, by combining versioned CAS with the program-order driver-follower barrier, \projecttitle{} guarantees preserving correctness without the latency penalties of global locks.

\section{\projecttitle{} Hardware Design}
\label{implementation}
\begin{figure}[t]
    \centering
    \includegraphics[width=\linewidth]{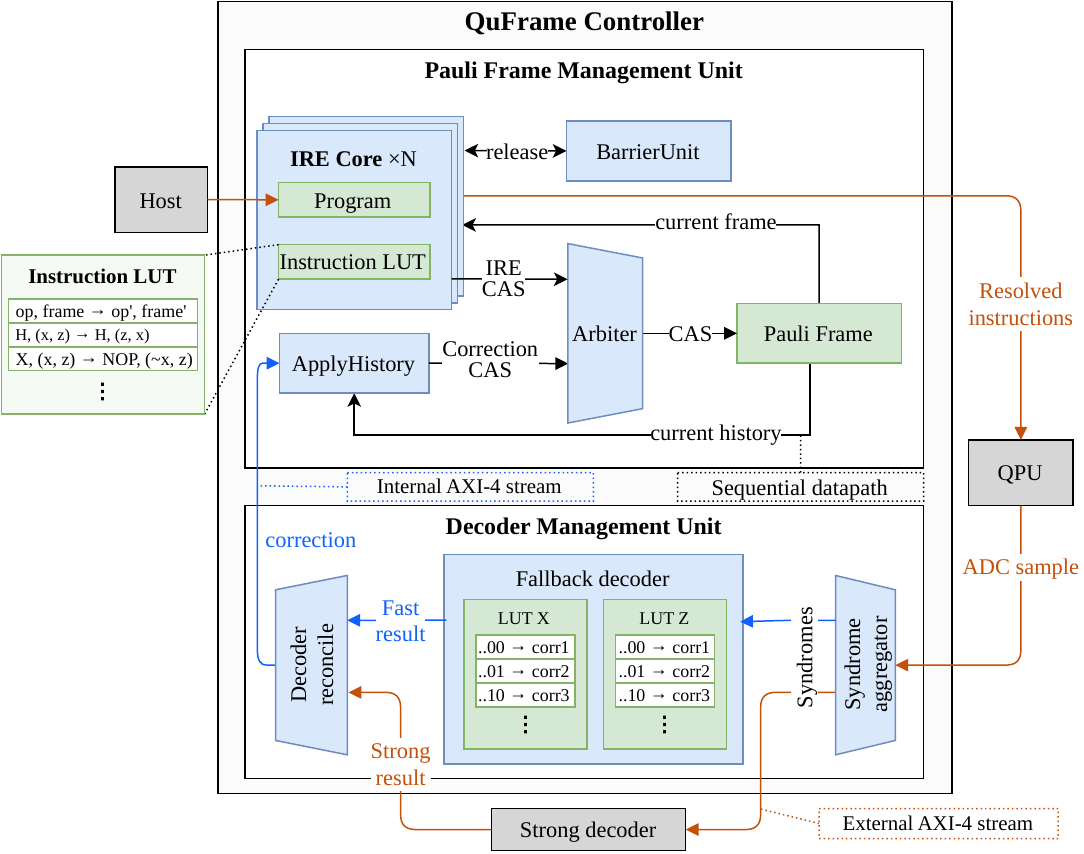}
    \caption{\projecttitle{} FPGA implementation. \emph{Grey boxes mark external elements, green boxes mark memory structures.}}
    \label{fig:fpga}
\end{figure}
\begin{figure*}[t]
    \centering
    \includegraphics[width=\linewidth]{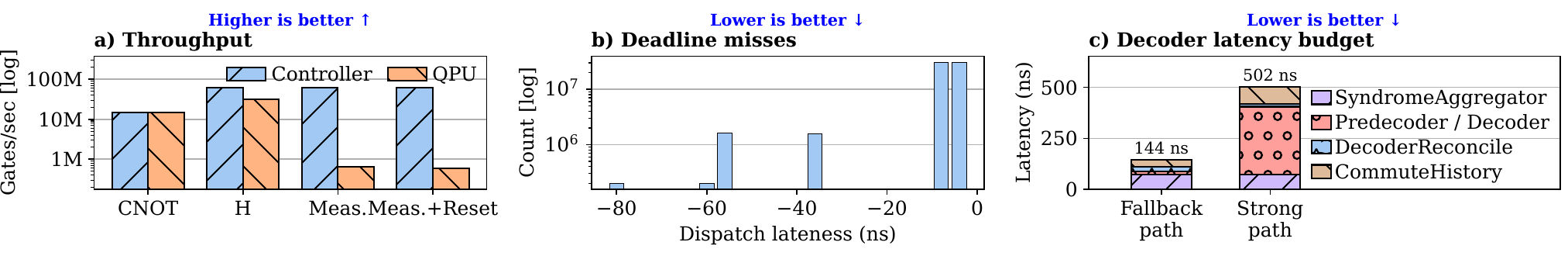}
    \caption{\projecttitle{} performance on the FPGA. \emph{a) Sustained gate throughput by gate type, against what the physical IBM Heron QPU itself can absorb. b) Dispatch lateness against each instruction's scheduled deadline. c) Decoder latency budget against the $1\,\mu\text{s}$ round.}}
    \label{fig:performance}
\end{figure*}

To enable real-time Pauli frame tracking and reconcile asynchronous decoding verdicts with nanosecond-scale instruction execution, \projecttitle{} separates control from the data plane, implementing a distributed, streaming hardware pipeline on an FPGA. 

\subsection{The \projecttitle{} Core}
\label{implementation:core}

The core forms the logical data plane, streaming operations through dedicated execution and decode units to guarantee deterministic runtime performance. Fig.~\ref{fig:fpga} locates the core within the real-time control workflow, showing its pipeline stages and external interfaces.

\myparagraph{System interfaces} The host streams independent instruction sequences per qubit to the controller over bidirectional 512-bit AXI4-Stream interfaces. A custom protocol splits execution into a \textit{load phase} (initializing program memory) and a \textit{run phase} (the Pauli Frame and Decoder Management Units executing the program and tracking corrections). We implement the syndrome provider, strong decoder, and simulated QPU as external modules that communicate via streams, mirroring a physical system setup. 

\myparagraph{Pipeline stages} Each qubit owns a dedicated IRE core with a three-stage pipeline: fetch, resolve, and commit. Cores fetch instructions from local SRAM, resolve them against the qubit's state via a combinational instruction LUT, and commit updates via CAS. Combinational resolution ensures constant latency, clearing uncontended instructions in three clock cycles. Failed commits incur a one-cycle penalty.


\myparagraph{Deterministic arbitration} Arbitration operates synchronously. A unified frame table processes all pending requests in one step per cycle, enforcing strict priority: decoder updates pre-empt two-qubit partner updates, which pre-empt local core updates. Unlike a global lock, resolution is combinational and only the single-cycle commit touches the table. Losing cores simply retry a failed CAS next cycle rather than stalling, keeping contention to bounded retries and avoiding serialized throughput. A hardware barrier synchronizes all cores at round boundaries.

\myparagraph{Decode dataflow} Measurement results flow through a non-blocking pipeline. An aggregator coalesces resolved samples into a per-round syndrome, which an on-chip fallback decoder resolves immediately. A reconciliation unit then compares it against the strong decoder's delayed response, emitting only the delta to the frame table. Bounded FIFOs decouple all stages.

\subsection{FPGA Architecture}
\label{implementation:fpga}

To meet strict timing constraints, we map the logical data plane of the \projecttitle{} core directly onto target FPGA primitives. 


\myparagraph{Deployment} We synthesize \projecttitle{} using Vitis HLS and deploy it on an AMD Alveo V80 FPGA as a single Coyote vFPGA kernel. Running at 250\,MHz, the 17-core prototype supports a distance-3 rotated surface code.

\myparagraph{Memory microarchitecture} We map datapath state structures directly to fixed-size registers. Each frame table row tracks two frame bits, a version tag, a timestamp, and a 16-entry history ring buffer. The rows are read from and written to in a static, parallel manner by the arbiter, which guarantees access correctness and allows the table to scale to more qubits.

\myparagraph{Logic and memory mapping} We map distinct lookup workloads to specific hardware resources. The instruction LUT encodes conjugation rules entirely in combinational logic to ensure deterministic latency. We implement the fallback decoder as read-only memory (ROM), which we pre-enumerate offline up to weight-6 and synthesize into LUTs. Runtime resolution therefore avoids the area overhead of online enumeration.

\section{Evaluation}
\label{sec:evaluation}

We evaluate \projecttitle{}'s overall performance and resource footprint (\S\ref{subsec:evaluation_performance}), followed by an ablation study of its core mechanisms (\S\ref{subsec:evaluation_ablation}).

\subsection{Methodology}
\label{subsec:evaluation_methodology}

We evaluate \projecttitle{} on an AMD Alveo V80 FPGA. The default baseline executes a rotated surface code (d=3) over 9 rounds across 50000 shots. We use d=3 because \projecttitle{} instantiates one IRE per physical qubit, so one patch already occupies 17 cores, already beyond the capacity of current controllers~\cite{guang2026breakingmemorybottlenecksquantum, Stefanazzi_2022}. The simulated QPU uses IBM Heron noise scaled by $10\times$ with $10\times$ longer decoherence times~\cite{AbuGhanem_2025, eccentric}. On-chip fallback decoding pairs with an external Helios decoder~\cite{helios} and a 16-entry history buffer. We note any setup variations.


\subsection{\projecttitle{} Performance and Hardware Cost}
\label{subsec:evaluation_performance}

\noindent\textbf{RQ1 (Logical error rate):} {\em To what extent does \projecttitle{} reduce the logical error rate compared to a serialized blocking controller?}

\myparagraph{Methodology} Because XQsim operates on a Pauli-based computation representation and lacks bare syndrome rounds~\cite{xqsim}, we compare \projecttitle{} against an ideal zero-latency simulation with an unbounded Helios decoder~\cite{helios} and a single-core baseline modeling the centralized Pauli frame unit and stall-on-decode pipeline shared by XQsim and prior proposals~\cite{riesebos2017, fu2019control, on2023}. We scale IBM Heron noise down by $5$--$20\times$ and evaluate both controllers on the FPGA.

\myparagraph{Results} As we present in Fig.~\ref{fig:controller_ler}, \projecttitle{} cuts the logical error rate by $38.5\%$ on average against the serialized blocking controller. The gap it closes relative to the zero-latency ideal widens across the sweep, rising from $42.3\%$ at $5\times$ to $65.5\%$ at $20\times$ (averaging $55.5\%$).

\statementLeftAccent{Takeaway \#1}{green}{\projecttitle{} cuts LER by $38.5\%$ on average against serialized blocking, closing up to $65.5\%$ of the zero-latency gap.}

\noindent\textbf{RQ2 (Dispatch throughput):} {\em What throughput does frame resolution sustain on the dispatch path?}

\myparagraph{Methodology} We instrument the dispatch time of every gate.

\myparagraph{Results} As we show in Fig.~\ref{fig:performance}a, \projecttitle{} sustains dispatch throughputs of 62.5\,M single-qubit and 15.3\,M two-qubit gates per second, against the 31.3\,M and 14.7\,M the QPU's 32\,ns and 68\,ns pulses can consume~\cite{AbuGhanem_2025}. Although two-qubit resolution can exceed the 68\,ns pulse, the scheduler masks it with concurrent readouts.

\statementLeftAccent{Takeaway \#2}{green}{Sustaining 15.3\,M two-qubit gates per second against the QPU's 14.7\,M, \projecttitle{} remains QPU-bound.}

\noindent\textbf{RQ3 (Deadline misses):} {\em How often does \projecttitle{} miss a deadline?}

\myparagraph{Methodology} We measure gate dispatch delay against scheduled execution deadlines under zero inter-gate latency.

\myparagraph{Results} Every dispatch meets its deadline (Fig.~\ref{fig:performance}b), keeping the QPU continuously busy. \projecttitle{} is bounded solely by QPU throughput, matching state-of-the-art controllers \cite{Stefanazzi_2022, qubic2023}.

\statementLeftAccent{Takeaway \#3}{green}{\projecttitle{} records zero deadline misses without inter-gate margin, sustaining the QPU throughput.}

\noindent\textbf{RQ4 (Decode path latency):} {\em What latency do the fallback and external decode paths incur relative to the $1$\,\textmu s round budget?}

\myparagraph{Methodology} We instrument both decode paths, timing each step for a representative round including external decoding.

\myparagraph{Results} The fallback path resolves in 144\,ns, half in syndrome aggregation (Fig.~\ref{fig:performance}c). The strong path requires 502\,ns, dominated by 333\,ns of algorithmic latency. Both finish within $1$\,\textmu s, but the $\approx 400$\,ns interconnect overhead (\S\ref{sec:motivation}) leaves the strong path marginal.

\statementLeftAccent{Takeaway \#4}{green}{\projecttitle{} completes the fallback path in 144\,ns, guaranteeing meeting the $1$\,\textmu s budget.}

\noindent\textbf{RQ5 (FPGA resources):} {\em What are \projecttitle{}'s resource requirements?}

\myparagraph{Methodology} We report the resource utilization of our hardware prototype across all FPGA primitives, excluding the strong decoder.

\myparagraph{Results} As we present in Table~\ref{tab:fpga_utilization}, the routed prototype consumes $11.06\%$ of LUTs and $6.60\%$ of flip-flops. Memories remain untouched: frame table, program memories, and decoder tables take $4.04\%$ of BRAM and $0.42\%$ of URAM. DSP usage is negligible at $1.88\%$. 

\statementLeftAccent{Takeaway \#5}{green}{\projecttitle{} leaves $88.94\%$ of the V80's LUTs and $98.12\%$ of its DSPs free for pulse generation.}


\begin{table}[t]
\centering
\caption{\projecttitle{} resource utilization on the AMD Alveo V80. \emph{Routed counts, with the share of the device in parentheses.}}
\label{tab:fpga_utilization}
\footnotesize
\setlength{\tabcolsep}{3pt}
\begin{tabular}{@{} c c c c c @{}}
\toprule
\textbf{LUT} & \textbf{FF} & \textbf{BRAM} & \textbf{DSP} & \textbf{URAM} \\
\midrule
284{,}604 (11.06\%) & 339{,}938 (6.60\%) & 151 (4.04\%) & 204 (1.88\%) & 8 (0.42\%) \\
\bottomrule
\end{tabular}
\end{table}

\begin{figure*}[t]
    \centering
    \includegraphics[width=\linewidth]{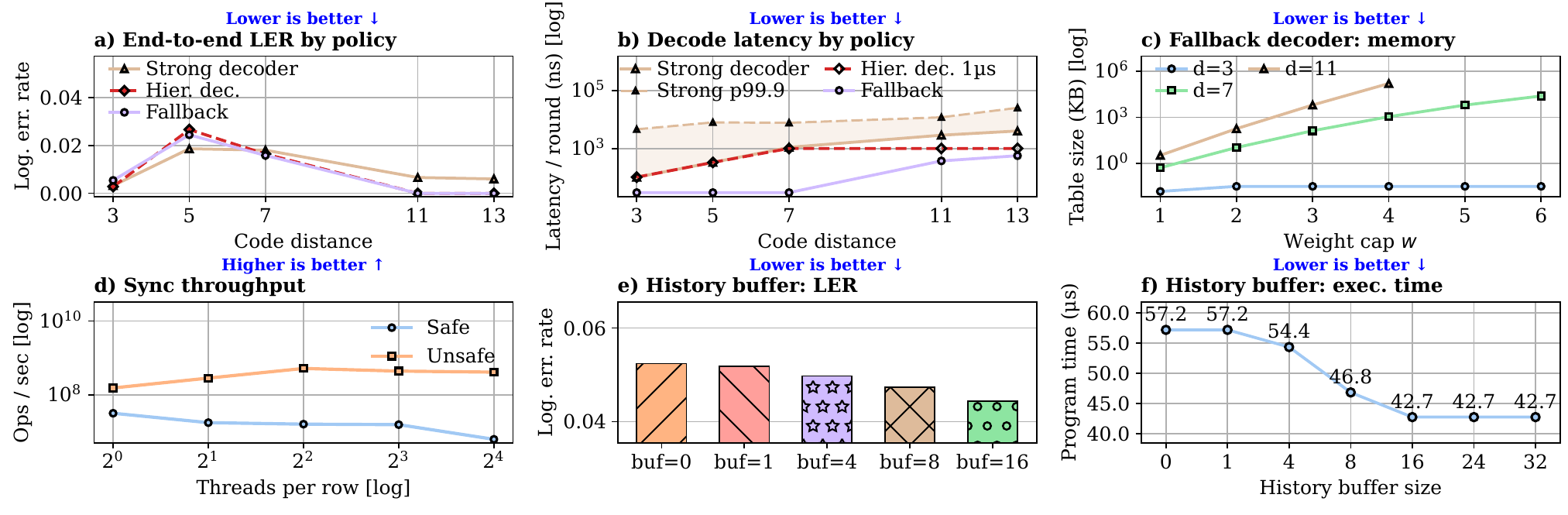}
    \caption{Decoder and synchronization ablations. \emph{a) End-to-end logical error rate of real \projecttitle{} runs. b) Decode latency, shading the strong decoder's median-to-$99.9$th-percentile spread. c) Fallback table size. d) Throughput vs.\ contending cores, with and without synchronization. e) Logical error rate vs.\ history buffer length. f) Program time vs.\ history buffer length.}}
    \label{fig:ablation}
\end{figure*}

\subsection{Ablation Study}
\label{subsec:evaluation_ablation}

\noindent\textbf{RQ6 (Decoding policies):} {\em How does the decoding hierarchy trade accuracy against the round budget?}

\myparagraph{Methodology}
We evaluate \projecttitle{} on rotated surface-code memory ($d$ rounds, $5{,}000$ shots) on a CPU host across three modes: \emph{fallback-only}, \emph{strong-only} (unbounded decode time), and \emph{hierarchical} (adopting the strong verdict if it resolves within $1$\,\textmu s, excluding link and commit latency), reporting latencies per round. We measure per-round decode latency in-process, while LER runs invoke the strong decoder out-of-process over a socket, capturing its round-trip overhead. Fallback LUT sizes are derived analytically by precomputing all syndromes with up to $w$ physical faults (error weight $w$).


\myparagraph{Results}
Hierarchical decoding lowers the LER from $1.05\%$ to $0.92\%$, as strong-only stalls accrue more idle decoherence than stronger corrections recover (Fig.~\ref{fig:ablation}a). Even under optimistic assumptions, strong-decoder latency scales poorly with distance: at $d=11$, it reaches $2.9$\,\textmu s at the median ($11.8$\,\textmu s at $p_{99.9}$), leaving only $0.8\%$ of decodes within the $1$\,\textmu s cutoff and none by $d=13$ (Fig.~\ref{fig:ablation}b). Tail latency violates the budget even at $d=3$ ($102$\,ns median), where $9.2\%$ of decodes miss the cutoff. In contrast, the fallback resolves in $30$\,ns through $d=7$ and $371$\,ns at $d=11$, where the table outgrows direct indexing. The fallback bounds per-round latency at the cost of LUT storage: $24$\,MB at $d=7$ ($w=6$), while capping $d=11$ at $w=3$ reduces its footprint from $153$\,MB ($w=4$) to $6.1$\,MB (Fig.~\ref{fig:ablation}c).

\statementLeftAccent{Takeaway \#6}{green}{Hierarchical decoding bounds latency at $1$\,\textmu s, while reducing mean error rate by $12.6\%$ at the cost of memory.}

\noindent\textbf{RQ7 (Synchronization cost):} {\em What impact does the synchronization mechanism have on throughput?}


\myparagraph{Methodology} We compare the versioned-CAS table against an unsynchronized one on a CPU host, sweeping 5--500 qubits and 1--16 threads per single row to stress contention.

\myparagraph{Results} Memory scales linearly at 3.2\,KB per qubit, while the unsynchronized table stays constant. Uncontended commits cost 32\,ns, a $4.8\times$ gap against the unsynchronized 152.6\,M ops/s (Fig.~\ref{fig:ablation}d). Contention latency reaches 162\,ns only under an extreme 16-core load.

\statementLeftAccent{Takeaway \#7}{green}{Synchronization adds just 3.2\,KB per qubit and 32\,ns per uncontended commit, which is negligible cost.}

\noindent\textbf{RQ8 (History buffer):} {\em How does history buffer capacity affect logical error rate and program time?}

\myparagraph{Methodology} Sweeping buffer depths on a transversal $H$ circuit ($d=3$, $18$ rounds) over $10^5$ shots, we measure expected execution time across $0$--$32$ and actual error rate across $0$--$16$ entries on the FPGA.


\myparagraph{Results} Expanding buffer capacity lowers the logical error rate by $15.4\%$ overall (Fig.~\ref{fig:ablation}e), with the largest marginal gain ($6.4\%$) from $8$ to $16$ entries. Eliminating gate stalls reduces execution time from $57.2$\,\textmu s (unbuffered) to $42.7$\,\textmu s at $16$ entries, plateauing through $32$ entries because the benchmark contains fewer than $16$ operations per qubit between rounds.

\statementLeftAccent{Takeaway \#8}{green}{A 16-entry buffer cuts the logical error rate by $15.4\%$ and program time by $25\%$, with no gain beyond 16 entries.}
\section{Related Work}
\label{sec:related_work}


\myparagraph{Hardware Pauli frame tracking}
Prior studies proposed integrating Pauli frame tracking into quantum control systems~\cite{riesebos2017, fu2019control, on2023}. However, these efforts remain theoretical, high-level abstractions that omit latency and throughput considerations. While XQsim~\cite{xqsim} synthesizes a control processor with a Pauli frame unit, its single lockstep pipeline blocks while awaiting corrections. Consequently, no existing design resolves Pauli frames across distributed execution datapaths without stalling.

\myparagraph{FPGA-based quantum controllers}
Existing platforms fall broadly into centralized single-core processors tailored to NISQ workloads~\cite{Stefanazzi_2022} and distributed systems that assign dedicated sequencer cores to individual qubits~\cite{qubic2023, fu2018micro, liu2025riscqgeneratorrealtimequantum, Yang_2022, Zhao_2025}. While distributed designs provide fine-grained pulse generation, none offer primitives for real-time, cross-core Pauli frame tracking and synchronization.

\myparagraph{Hierarchical QEC decoding}
Hierarchical decoding reconciles accuracy with throughput by structuring solvers into multi-tiered pipelines. Existing strategies typically pair a low-latency heuristic or table-driven lookup with a high-fidelity engine via weight-based fallbacks~\cite{toshio2025decoderswitchingbreakingspeedaccuracy, basak2026hierarchicalquantumdecoders, lilliput}, cascade predecoders to sparsify syndrome streams~\cite{delfosse2020hierarchicaldecodingreducehardware, Knapen_2026, wegmann2026zerogpredecoderawaredecoderquantum}, or speculate across parallel decoding windows and re-decode on boundary mispredictions~\cite{swiper}. While \projecttitle{} incorporates a two-tier fallback model, it decouples the decoding algorithm from a non-blocking controller interface that reconciles speculative Pauli frame updates in hardware for strict real-time guarantees.

\statementLeftAccent{How \projecttitle{} differs}{blue}{\projecttitle{} is the first microarchitecture
to track Pauli frames at line rate across concurrent cores and
multi-tier decoders under strict real-time deadlines.}

\section{Conclusion}
\label{conclusion}


Real-time error correction is vital for non-Clifford execution: magic-state gadget measurements resolve correctly only if the controller tracks preceding Pauli corrections without latency penalties. \projecttitle{} accomplishes this at datapath speeds, elevating per-qubit Pauli frame tracking to a first-class, code- and decoder-agnostic microarchitectural primitive. QuFrameISA decouples frame tracking from control flow, enabling distributed Instruction Resolution Engines to resolve gates across arbitrary QEC codes via single table lookups. Concurrently, a lock-free Pauli Frame Table eliminates cross-lane serialization, while an on-chip fallback decoder with asynchronous reconciliation meets the strict $1$\,\textmu s cycle deadline for any plug-in decoder without stalling execution. Together, these mechanisms sustain full throughput with no inter-gate margin.


\bibliographystyle{ACM-Reference-Format}
\bibliography{bibliography}

\end{document}